\documentclass[
 reprint,
superscriptaddress,
 amsmath,amssymb,  
 aps,
]{revtex4-2}
\usepackage[utf8]{inputenc}
\usepackage[T1]{fontenc}
\usepackage{lineno}
\usepackage{graphicx}
\usepackage{xcolor}
\usepackage{svg}
\usepackage[hypertexnames=false]{hyperref}
\usepackage{float}
\usepackage{braket}
\usepackage{siunitx}
\preprint{APS/123-QED}
\usepackage{algorithm}
\usepackage{algpseudocode}
\hypersetup{colorlinks=true,
linkcolor=blue,
citecolor=blue,
urlcolor=blue,
filecolor=blue
}
\usepackage{dsfont}
\usepackage{physics}
\usepackage{flushend}
\usepackage{mathtools}
\usepackage{bbold}
\usepackage{bm}
\usepackage{color} 
\usepackage{tcolorbox}
\usepackage{soul}

\newcommand{\beq}{\begin{equation}}
\newcommand{\eeq}{\end{equation}}

\DeclareRobustCommand{\SkipTocEntry}[5]{}

\newcommand{\EVS}{\ensuremath{E_\mathrm{VS}}} 
\newcommand{\EST}{\ensuremath{E_\mathrm{ST}}} 

\begin{document}
\title{Mapping of valley-splitting by conveyor-mode spin-coherent electron shuttling}

\author{Mats Volmer}
\thanks{These authors contributed equally.}
\affiliation{JARA-FIT Institute for Quantum Information, Forschungszentrum J\"ulich GmbH and RWTH Aachen University, Aachen, Germany}
\author{Tom Struck}
\thanks{These authors contributed equally.}
\affiliation{JARA-FIT Institute for Quantum Information, Forschungszentrum J\"ulich GmbH and RWTH Aachen University, Aachen, Germany}
\affiliation{ARQUE Systems GmbH, 52074 Aachen, Germany}
\author{Arnau Sala}
\author{Bingjie Chen}
\author{Max Oberländer}
\author{Tobias Offermann}
\author{Ran Xue}
\author{Lino Visser}
\affiliation{JARA-FIT Institute for Quantum Information, Forschungszentrum J\"ulich GmbH and RWTH Aachen University, Aachen, Germany}
\author{Jhih-Sian Tu}
\author{Stefan Trellenkamp}
\affiliation{Helmholtz Nano Facility (HNF), Forschungszentrum J\"ulich, J\"ulich, Germany}
\author{{\L}ukasz Cywi{\'n}ski}
\affiliation{Institute of Physics, Polish Academy of Sciences, Warsaw, Poland}
\author{Hendrik Bluhm}
\author{Lars R. Schreiber}
\email{lars.schreiber@physik.rwth-aachen.de}
\affiliation{JARA-FIT Institute for Quantum Information, Forschungszentrum J\"ulich GmbH and RWTH Aachen University, Aachen, Germany}
\affiliation{ARQUE Systems GmbH, 52074 Aachen, Germany}

\begin{abstract}
In Si/SiGe heterostructures, the low-lying excited valley state seriously limit operability and scalability of electron spin qubits. For characterizing and understanding the local variations in valley splitting, fast probing methods with high spatial and energy resolution are lacking. Leveraging the spatial control granted by conveyor-mode spin-coherent electron shuttling, we introduce a method for two-dimensional mapping of the local valley splitting by detecting magnetic field dependent anticrossings of ground and excited valley states using entangled electron spin-pairs as a probe. The method has sub-$\si{\micro \electronvolt}$ energy accuracy and a nanometer lateral resolution. The histogram of valley splittings spanning a large area of 210$\,$nm by 18$\,$nm matches well with statistics obtained by the established but time-consuming magnetospectroscopy method. For the specific heterostructure, we find a nearly Gaussian distribution of valley splittings and a correlation length similar to the quantum dot size. Our mapping method may become a valuable tool for engineering Si/SiGe heterostructures for scalable quantum computing. 
\end{abstract}

\flushbottom
\maketitle
\addtocontents{toc}{\SkipTocEntry}
\section*{Introduction}
Si/SiGe-heterostructures are one of the most promising host materials for spin qubits~\cite{Stano22}, as they offer low potential fluctuations, charge noise, long coherence times~\cite{Struck2020}, high-fidelity control~\cite{Yoneda2018, Xue2022, Noiri2022, Mills2022} and are industry-compatible platforms that allow for fabrication in established silicon production lines~\cite{Neyens23}. However, some devices exhibit low lying valley states that limit high-temperature operation of spin-initialization, -manipulation and Pauli-spin blockade readout, and hinder spin-shuttling \cite{Kawakami2014, Vandersypen17, Ferdous2018, Langrock23}. Local minima in the energy splitting between the low-lying valley states, \EVS{},  pose the main obstacle for the scalability of this platform. Innovations in growth and fabrication strategies~\cite{Losert23, Woods23, Neul23}, but also efficient methods to benchmark the local valley splitting are needed to overcome it. 

A large range of local \EVS{}, from \SI{6}{\micro \electronvolt} to more than \SI{200}{\micro \electronvolt}, was observed in gate-defined quantum dots (QDs) formed in Si/SiGe-heterostructures~\cite{Borselli11, Shi2011, Kawakami2014, Zajac2015, Mi17, Jones2019, Borjans2019, Hollmann20, McJunkin21, Chen21, Dodson22, Denisov23, Esposti23}. The \EVS{} is theorized to be a randomly distributed local material parameter, subject to atomic-scale crystal variations \cite{Friesen2007, Friesen2010, Culcer2010, Hosseinkhani2020, Wuetz2022, Losert23} of the Si/SiGe-heterostructure. Thus a few measurements of \EVS{} at different spots do not suffice to confidently benchmark the quality of a heterostructure \cite{Wuetz2022}. Many different methods to determine the \EVS{} of a Si/SiGe QD were reported, such as thermal excitation \cite{Kawakami2014}, pulsed-gate spectroscopy in a single \cite{Hollmann20,  Dodson22} or double \cite{Chen21} QD and the identification of the spin-valley relaxation hot-spot \cite{Borjans2019, Hollmann20}. Other methods measure the singlet-triplet energy splitting \EST{}, being a lower bound of the \EVS{}, by Pauli-spin blockade \cite{Jones2019, Wuetz2022} or magnetospectroscopy \cite{Borselli11, Shi2011, Zajac2015, McJunkin21, Dodson22, Esposti23}. High-energy resolution has been achieved by dispersive coupling to a resonator \cite{Burkhard2016, Mi17}, and some attempts towards laterally mapping \EVS{} \cite{Hollmann20, Dodson22, Denisov23} have been published, but these are involved, time-consuming and cover a small area. Determining \EVS{} by Shubnikov-de-Haas oscillations \cite{Mi2015} lacks lateral resolution and tends to overestimate \EVS{} due to localization by the out-of-plane magnetic field \cite{Friesen2006}. To this end, we need a time-efficient method with good energy resolution that can map the valley splitting landscape of a realistic Si/SiGe quantum chip. 

In this work, we present an efficient method for mapping the local valley splitting in silicon across a large area with a resolution that can capture the local variations of \EVS{}. We employ singlet-triplet oscillations of a spatially separated pair of spin-entangled electrons, with one of them shuttled to a distant position as a probe to locally detect magnetic field-induced anticrossings between spin-valley states, from which we then obtain a magnitude for \EVS{} \cite{Jock22}. Leveraging coherent conveyor-mode shuttling~\cite{Langrock23, Seidler22,Xue23,Struck23}, we extend this analysis to create a dense one-dimensional map of the valley splitting for a Spin-Qubit-Shuttle (SQS)~\cite{Langrock23,Struck23,Note1}.
Our method yields a nanometer-resolution along the shuttle direction, which suffices to resolve local features in the valley splitting landscape depending on the QD size. 
By applying voltage offsets to two long gates parallel to the shuttle direction, the shuttle trajectory can be displaced (here up to \SI{18}{\nano \meter}), which result in a two-dimensional map of \EVS{}. We thus present four valley splitting traces, each with an approximate length of \SI{210}{\nano \meter}, with 150 \EVS{} measurements per trace and a sub-\SI{}{\micro \electronvolt} energy uncertainty. We report measured values of the valley splitting that range from \SI{4.6}{\micro \electronvolt} to \SI{59.9}{\micro \electronvolt}, and that exhibit a continuous behavior punctuated by sudden jumps. We attribute these rapid changes to unintentional tunneling events during conveyor-mode shuttling, which we can mitigate by displacing the channel vertically. Our method enables efficient valley splitting mapping, which provides sufficient statistics to infer an accurate mean and shape of the distribution by single electron spin shuttling.

\begin{figure}
    \centering
    \includegraphics[width=\linewidth]{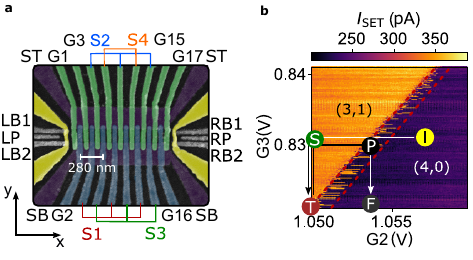}
    \caption{Spin Qubit Shuttle (SQS) device and experimental method. (a) False-colored scanning electron micrograph (SEM) of the device used in the experiment, showing a top-view on the three metallic layers (1st purple, 2nd blue, 3rd green) of the SQS, and their electrical connection scheme. At both ends single-electron transistors (SETs) are formed in the quantum well by gates LB1, LB2, and LP (RB1, RB2, and RP, respectively) on the second gate layer, with the current path induced by the yellow gates on the third layer. (b) Charge stability diagram of the outermost left DQD recorded by the left SET current $I_{\mathrm{SET}}$. DQD fillings are indicated by $(n,m)$, with \(n\) and \(m\) denoting the number of electrons in the left and right QDs, respectively. The red dashed lines indicate boundaries of the PSB region. Labelled circles indicate voltages on G2 and G3 and correspond to pulse stages used in subsequent experiments. Arrows indicate pulse order. Pulse stages T as well as F reach down to $V_\mathrm{G3}=\SI{0.7}{\volt}$.}
    \label{fig:overview}
\end{figure}
\addtocontents{toc}{\SkipTocEntry}
\subsection*{Device Layout}
The device used for the experiments is the same as that described in Ref.~\cite{Struck23}. It comprises three Ti/Pt gate layers, separated by \SI{7.7}{\nano\meter} thick Al$_2$O$_3$, and is fabricated on an undoped Si/Si$_{0.7}$Ge$_{0.3}$ quantum well (see method section for layer stack). The one-dimensional electron channel (1DEC) is formed by an approximately \SI{1.2}{\micro\meter}-long split-gate with \SI{200}{\nano\meter} spacing (shown in purple in Fig.~\ref{fig:overview}a). By applying DC voltages $V_\mathrm{ST}, V_\mathrm{SB}$ to the split-gate, the 1DEC is confined in y-direction. Seventeen clavier gates are fabricated on top of the device, with a combined gate pitch of \SI{70}{\nano\meter}. Of these, eight are on the second metal-layer and labelled G2, G16, 3$\times$S1, and 3$\times$S3, while nine are on the third metal-layer and labelled G1, G3, G15, G17, 3$\times$S2, and 2$\times$S4. In conveyor mode~\cite{Langrock23}, two to three clavier gates are electrically connected to four so-called shuttle gates S1, S2, S3, and S4 \cite{Xue23,Struck23,Kuenne23}. The shuttle gates are named differently, as each shuttle gate comprises more than one clavier gate as indicated in Fig. \ref{fig:overview}a. As a result, every fourth clavier gate shares the same potential, which leads to a periodic electrostatic potential with a period of $\lambda = 280$\,nm. Generating a travelling wave potential (see methods section for details on electron shuttling in conveyor-mode), we coherently shuttle the electron spin for a nominal distance of up to \SI{336}{\nano \meter} in a global in-plane magnetic field $B$. We shuttle at a frequency of \SI{10}{\mega \hertz} which corresponds to an electron velocity of \SI{2.8}{\meter \per \second}. The SQS has a single electron transistor (SET) at each end, serving as electron reservoirs and proximity charge sensors.

\addtocontents{toc}{\SkipTocEntry} 
\section*{Results}
\addtocontents{toc}{\SkipTocEntry}
\subsection*{DQD Valley Splitting Measurement}
As a basis for the \EVS{} mapping technique discussed later, we first consider a method to determine \EVS{} in a static double quantum dot (DQD). Therefore, next to the left SET, we form a DQD under gate G2 and the leftmost clavier gate from S1. Gates G1, G3 and the leftmost clavier gate of S2 act as barrier gates. Fig.~\ref{fig:overview}b displays a charge stability diagram for the DQD. We measure the valley splittings $E_{l}$ ($E_{r}$) of the left (right) QD of the DQD using singlet-triplet oscillations, which probe the magnetic anticrossings induced by spin-valley couplings in each QD.

\begin{figure}
    \centering
    \includegraphics[width=\linewidth]{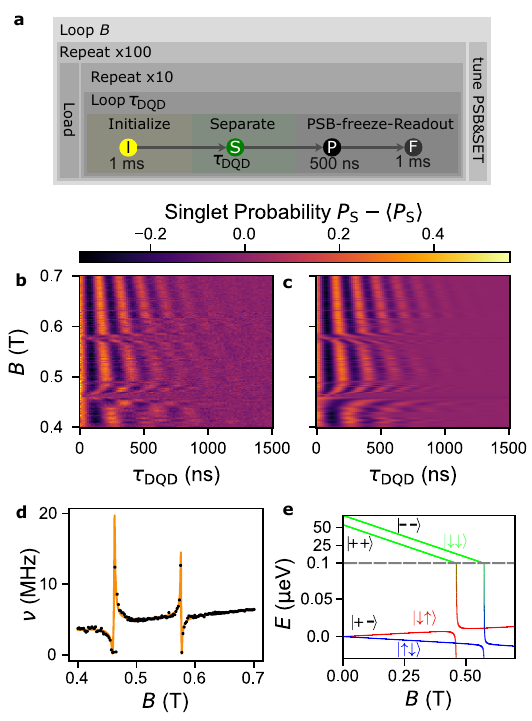}
    \caption{Spin-valley anticrossing in a DQD. (a) Experiment flowchart explaining the microscopic pulse stages, parameter loops as well as stabilizing measures. Waiting times at pulse stages are indicated by times below. (b) Normalized singlet return probability  as a function of the magnetic field $B$ and DQD separation time $\tau_\text{DQD}$. The singlet return-probability $P_\mathrm{S}$ is normalized such that each horizontal line averages to zero. (c) Fit to the data from (b) using Eq.~(\ref{eq:gauss}). (d) Frequencies extracted from the fit in (c). The orange curve is a least-square fit to the data. Uncertainties of frequencies are on the order of \SI{100}{\kilo \hertz} and smaller than the size of the black dots. (e) Energy spectrum of the Hamiltonian from Eq.~(\ref{eq:ham}). The color mixture represents the spin state composed from  colors of labeled spin base state, while the black symbols label the valley state.
    For clarity, the energy axis is upscaled around the $\ket{\uparrow\downarrow +-}$ and $\ket{\downarrow\uparrow +-}$ states with spin projection along the $z$ axis $m_\mathrm{S}=0$. For these states, their magnetic field dependence, proportional to $\Delta g \mu_B$, is four orders of magnitude smaller than that of the states $\ket{\downarrow\downarrow --}$ and $\ket{\downarrow\downarrow ++}$, with $m_\mathrm{S}=-1$. The parameters used in (e) are extracted from the fit in (d).}
    \label{fig:Anticrossing}
\end{figure}

To this end, we apply the following pulse sequence: We load four electrons into the leftmost QD for $1\,$ms to initialize into a spin-singlet (S) state in the (4,0) charge state~\cite{Struck23} (Fig.~\ref{fig:overview}b, stage I). Next, we split the spin-singlet by rapidly pulsing to the (3,1) charge state (stages I $\to$ S) within a rise-time of $\approx\SI{1.2}{\nano \second}$ (limited by \SI{300}{\mega \hertz} bandwidth of our waveform generator). As a function of wait-time $\tau_\text{DQD}$, singlet-triplet oscillations occur with a frequency $\nu$ proportional to $B$, and the difference of the electron g-factors $\Delta g$ of the DQD. For detection of the S-state, we pulse into the Pauli-Spin-blockade (PSB) (area between red dashed lines in Fig.~\ref{fig:overview}b) and wait for $500\,$ns. The PSB charge state is read out by the SET current $I_\mathrm{SET}$, after freezing this charge state by reducing the DQD tunnel-coupling (stages P $\to$ F; $V_\mathrm{G3}(\mathrm{F})\approx\SI{0.7}{\volt}$)~\cite{Nurizzo23}. There, we read the charge state via measuring $I_\mathrm{SET}$ for \SI{1}{\milli \second}. We repeat this pulse sequence (Fig.~\ref{fig:Anticrossing}a) while varying $\tau_\text{DQD}$ from 0 to $1.5\,$\si{\micro \second}, in 100 equidistant time steps. Repeating this loop 1000 times, we calculate the spin-singlet return probability $P_\mathrm{S}(\tau_\text{DQD})$ at a set $B$ (Fig.~\ref{fig:Anticrossing}b), while every 10 loop-iterations the correct electron filling of the DQD is reinitialized as a precaution. In order to counter slow noise related drifts on the PSB and the SET, both the PSB-stage voltage as well as the SET voltages are retuned after 1000 loop iterations (details in methods section).

The singlet-triplet oscillation frequency $\nu$ contains the important information and is extracted as follows. We fit the measured $P_\mathrm{S}(\tau_\text{DQD},B)$  line by line to 
\begin{equation}\label{eq:gauss}
P_\mathrm{S}(\tau_\mathrm{DQD})=a \exp\!\left(-\frac{\tau_\mathrm{DQD}^2}{{T_2^*}^2}\right) \cos\!\left(2\pi \nu \tau_\mathrm{DQD} +\varphi\right)+c,
\end{equation}
\noindent where $a$, $\nu$, $\varphi$ are the visibility, frequency and phase of the spin-singlet-triplet oscillations, respectively, and $T_2^*$ is the ensemble spin-dephasing time of the entangled spin-state. The offset $c$ is partly absorbed by subtraction of the linewise mean $\left< P_\mathrm{S}(\tau_\mathrm{DQD}) \right>$. The fit with a Gaussian decay (Fig.~\ref{fig:Anticrossing}c) captures all the relevant features of the measured data (cf. Fig.~\ref{fig:Anticrossing}b). Here, we are interested in $\nu(B)$ (black dots in (Fig.~\ref{fig:Anticrossing}d)), which reveals two distinct anticrossings on top of a constant slope $p$. The slope is expected to be proportional to $\Delta g=\frac{p h}{\mu_B}$ (with $h$ and $\mu_B$ Planck's constant and Bohr-magneton, respectively) provided the effective magnetic field gradient due to $\Delta g$ exceeds the Overhauser field gradient ($\sim$0.01\,mT) of the randomly fluctuating $^{29}$Si and $^{73}$Ge nuclear spin-baths. As this condition is easily fulfilled, we can fit $\Delta g$ (Tab.~\ref{tab:ac_fit_params}). 

Next, we argue that the two anticrossings stem from the spin-valley coupling in each of the QDs, and can be employed as a precise probe for the valley splittings $E_{l}$ and $E_{r}$. As we will show in the following sections, this anticrossing is crucial for mapping the valley splitting by coherent spin shuttling. 
We assume that intervalley tunneling couples higher energy valley, $\ket{+}$, in the left QD to the lower energy valley, $\ket{-}$ in the right QD, so that charge separation  $(4,0)\! \rightarrow (3,1)$ creates a state in which two electrons form a spin singlet in the $\ket{-}$ valley in the left QD, and thus are inert \cite{Connors22, Cai23}, while the remaining two electrons form a spin singlet involving $\ket{+}$ valley in the left QD and $\ket{-}$ valley in the right QD.
Deep in the $(3,1)$ regime the dynamics in the relevant space of four lowest-energy states is modeled with a Hamiltonian
\begin{align}\label{eq:ham}
    H = \left(\begin{array}{cccc}
    -\Delta E_\text{Z}/2 & 0 & 0 & v_l \\
    0 & \Delta E_\text{Z}/2 & v_r & 0 \\
    0 & v_r & E_{r}-\bar{E}_{\text{Z},+}& 0\\
    v_l & 0 & 0 & E_{l}-\bar{E}_{\text{Z},-}
    \end{array}
    \right)\! 
\end{align}
written in the basis of $\left\{\ket{\uparrow\downarrow +-},\ket{\downarrow\uparrow +-},\ket{\downarrow\downarrow ++},\ket{\downarrow\downarrow --}\right\}$, where the first (second) arrow and sign indicate the spin and valley state of the non-inert electrons in the left (right) QD. Note that in the $\ket{\downarrow\downarrow --}$ state the two inert electrons are in the $\ket{+}$ valleys state of the left QD. $\Delta E_\text{Z} = \mu_B B(g_{r,-} - g_{l,+})$ is the difference between the Zeeman energies of the two electrons with opposite spin in different valley states, which results from the $g$-factor difference between an electron in the right QD and $\ket{-}$ valley (with $g$-factor $g_{r,-}$) and an electron in the left QD and $\ket{+}$ valley (with $g$-factor $g_{l,+}$). $\bar{E}_{\text{Z},+}$ ($\bar{E}_{\text{Z},-}$) is the Zeeman energy for two electrons with parallel spins in the $\ket{++}$ ($\ket{--}$) state. Fits of data with a model involving also $(4,0)$ state, and tunnel coupling, $t_c$ in the DQD, confirmed that $t_c$ has negligible effect on spin dynamics in $(3,1)$ regime. As explained above, the Overhauser field is disregarded. 

We diagonalize the Hamiltonian and fit $\nu(B)$ in Fig.~\ref{fig:Anticrossing}d (orange line) with parameters shown in Tab.~\ref{tab:ac_fit_params} corresponding to the energy spectrum shown in Fig.~\ref{fig:Anticrossing}e. Note that the assignment of the anticrossings to the left and right QDs is arbitrary at this stage of the analysis; the indices $l$ and $r$ in Tab.~\ref{tab:ac_fit_params} can be swapped. Our model fits $\nu(B)$ very well. Hence, the occurrence of spin-valley anticrossings does not require any tunnel-coupling in the DQD except from initialization and detection of the S-state. This notion is decisive for valley-mapping by shuttling, which involves separation of the two electrons. Assignment for the valley-splitting is straightforward: the magnetic field $B_\mathrm{VS}$ in the center of the anticrossing can be converted to a \EVS{} by $B_\mathrm{VS}=\EVS{} \mu_B/g$, where $g=2$ and the width of the anticrossing is proportional to the coupling strength $v$. A similar analysis of a DQD formed at different screening gate voltages can be found in supplementary Fig.~S1.

\begin{table}[t]
    \centering
    \caption{Fit parameters, together with their uncertainty, for the model presented in Eq.~(\ref{eq:ham}), using the data from Fig.~\ref{fig:Anticrossing}c and e. For the coupling elements $v_r$ and $v_l$, we also indicate the states that are coupled.}
    \begin{tabular}{|c|c|c|c|c|}
    \hline
        Parameter & value & $1 \sigma$ & unit/factor & coupling states \\
    \hline
    \hline
       $\Delta g$  & 6.58  & 0.04 &  $ 10^{-4}$ & - \\
    \hline         
       $E_{r}$  & 53.52 & 0.17 & \si{\micro \electronvolt}& - \\
    \hline         
       $E_{l}$  & 66.64 & 0.04 & \si{\micro \electronvolt}& - \\
    \hline
        $v_r$  & 82 & 14 & neV & $\ket{\downarrow\uparrow +-}$ \& $\ket{\downarrow\downarrow ++}$ \\
    \hline
        $v_l$  & 58 & 3 & neV & $\ket{\uparrow\downarrow +-}$ \& $\ket{\downarrow\downarrow --}$ \\
    \hline
    \end{tabular}
    \label{tab:ac_fit_params}
\end{table}

\addtocontents{toc}{\SkipTocEntry}
\subsection*{Valley splitting mapping}

\begin{figure}
    \centering
    \includegraphics[width=\linewidth]{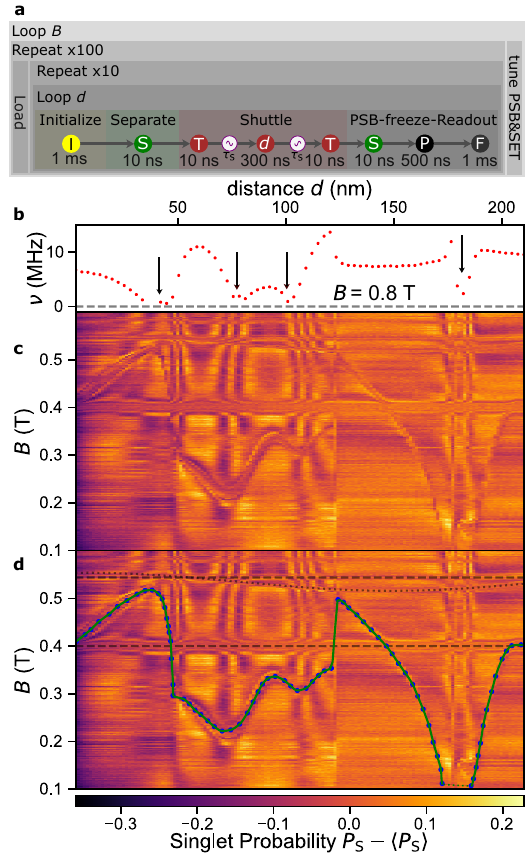}
    \caption{Mapping of the local valley splitting using the ST$_0$ oscillations. (a) Flowchart of the microscopic pulse stages, parameter loops as well as stabilizing measures. Waiting times at pulse stages are indicated below. Compared to Fig.~\ref{fig:Anticrossing}a, the electron is shuttled by a distance $d$, waits there for $\tau_W=\SI{300}{\nano \second}$ and is shuttled back, prior to PSB. (b) Extracted frequencies $\nu(d)$ measured at a magnetic field of \SI{0.8}{\tesla}. The $1\sigma$-intervals are smaller than the symbols. (c) Raw data of the singlet return probability $P_S$ as a function of shuttle distance $d$ and magnetic field $B$. To enhance contrast, we subtract the averaged return probability $\langle P_S \rangle$ for each $B$. (d) same as panel (c) with additional markers (see text). The spin-valley anticrossing of the shuttled QD is indicated by blue points connected by a green spline curve.}
    \label{fig:Mapping}
\end{figure}

Next, we discuss the use of the spin-valley anticrossing in a QD for mapping the valley splitting along the 1DEC. Therefore, in addition to the pulse scheme explained above (Fig.~\ref{fig:Anticrossing}a), we shuttle the electron-spin in the right QD fast by a distance $d(\tau_\mathrm{S})$ (for shuttle time $\tau_\mathrm{S}$, see Eq.~(\ref{eq: Shuttle pulse}) in the method section), let the entangled singlet-triplet-state evolve for a fixed waiting period ($\tau_\mathrm{w}=\SI{300}{\nano \second}$) and then shuttle it back by the same distance for PSB detection. Thus, the pulse scheme for mapping (Fig.~\ref{fig:Mapping}a) is complemented by the 10\,ns long stage T (voltages in Fig.~\ref{fig:overview}b), a shuttle pulse for time $\tau_\mathrm{S}$, a fixed waiting period at stage d, the time reversed shuttle pulse to enter stage T (DQD with large barrier) followed by stage S, the detuned tunnel-coupled DQD in charge state (3,1). Note that compared to the pulse scheme (Fig.~\ref{fig:Anticrossing}a), we measure $P_\mathrm{S}(d,B)$ instead of $P_\mathrm{S}(\tau_\text{DQD},B)$, which turns out to be sufficient for mapping the valley splitting. Another parameter that can be varied is $\tau_\mathrm{w}$ in stage d. Measurements of the three-dimensional parameter space $P_\mathrm{S}(d,\tau_\mathrm{w}, B)$ are shown in supplementary Fig.~S2. A scan $P_\mathrm{S}(d,\tau_\mathrm{w}, B=800\,$mT$)$ is employed to probe $\nu(d)$ fitted by Eq.~(\ref{eq:gauss}) with $\tau_\mathrm{w}$ replacing $\tau_\text{DQD}$ (Fig.~\ref{fig:Mapping}b). Notably, the fitted frequency of the singlet-triplet oscillations $\nu(d)$ varies smoothly, with exception at $d\approx 120\,$nm, and drops close to zero at some $d_i$ (black arrows). Presumably, $\nu(d)$ is governed mainly by variations of the electron g-factor in the propagating QD due to variations in confinement. 
These are expected partly due to the deterministic breathing of the confinement potential of the moving QD, partly due to electrostatic disorder in the quantum well \cite{Langrock23}. Note that we cannot distinguish by measurement of $\nu(d)$, which of the QDs has the larger electron g-factor.  

The local variations of the $g$-factor difference helps us to understand features in $P_\mathrm{S}(d,B)$ (Fig.~\ref{fig:Mapping}c), our main result. Curved (spaghetti-like) features clearly visible on top of background that appear when changes of $P_S(d,B)$ along a certain direction in the $(d,B)$ plane are much larger than changes along the corresponding perpendicular direction. For example, at distances $d_i$ (highlighted by arrows in Fig.~\ref{fig:Mapping}b), at which $\nu(d)$ approaches zero, the $P_S$ signal weakly depends on $B$, while it depends strongly on $d$ (due to strong variation of $\nu(d)$, see Fig.~\ref{fig:Mapping}b), resulting in appearance of vertical features. Besides some horizontal features (marked by black dashed lines in Fig.~\ref{fig:Mapping}d), which we explain below, there is a continuous widely varying feature marked by the green solid line in Fig.~\ref{fig:Mapping}d (details in supplementary section S5). This line follows the spin-valley anticrossing of the shuttled electron spin. It is generated by waiting at $d$ for $\tau_\mathrm{w}=\SI{300}{\nano \second}$ and accumulating phase due to a relatively large modification of the  singlet-triplet oscillation frequency at the anticrossing. It is thus a measure of $\EVS{}(d)$ along the 1DEC. We support this notion by the $P_\mathrm{S}(d,\tau_\mathrm{w}, B)$ data shown in supplementary section S4.

Notably, at $d=0$\,nm and $B\approx \SI{0.4}{\tesla}$, this line overlaps with a horizontal feature (marked by the lower dashed line in Fig.~\ref{fig:Mapping}d) and the B-field matches with one of the \EVS{} of the DQD. This $d$-independent feature originates from the accumulation of a phase during the stages S and T, at which the DQD in charge state (3,1) is formed. There, the total waiting period is $\SI{40}{\nano \second}$ (Fig.~\ref{fig:Mapping}a), which is sufficient to identify the anticrossing by the singlet-triplet oscillations (cf. Fig.~\ref{fig:Anticrossing}b). Presumably, this horizontal line is broadened in $B$ as the QD position is slightly displaced in stage T compared to stage S, altering the $B$ at which the anticrossing occurs. Now, it is justified to attribute this anticrossing to the right QD. The index of $E_{r}$ in Tab. \ref{tab:ac_fit_params} is therefore correct.

The counterpart of the lower horizontal line is the upper horizontal line at $B=\SI{0.54}{\tesla}$, which matches $E_{l}$ in Tab. \ref{tab:ac_fit_params}. At its origin ($d=0$\,nm), a wavy feature (black dotted line in Fig.~\ref{fig:Mapping}d) around the upper dashed line is barely visible. We assign this line to the spin-valley anticrossing of the left (static) QD, due to which a phase is accumulated during $\tau_\mathrm{w}=\SI{300}{\nano \second}$. This is expected, since the sinusoidal voltages applied to the shuttle gates capacitivly cross-couple to the left QD. Hence, the left QD is slightly displaced by the same period as the period of the shuttle voltages, and thus its valley-splitting gets a tiny $d$-dependence with this period. This matches exactly the observation in Fig.~\ref{fig:Mapping}c,d.

\begin{figure}
    \centering
    \includegraphics[width=\linewidth]{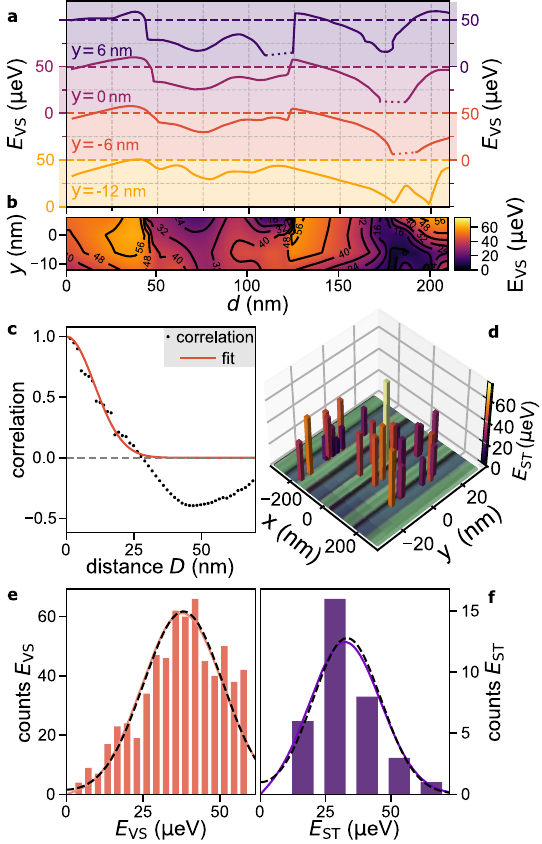}
    \caption{Comparison of valley splitting mapping techniques. (a) Four \EVS{} scan-lines of valley splitting along different $y$-displacements measured by the same method as the data shown in Fig.~\ref{fig:Mapping} with the curve at $y=0$\,nm being taken from its panel d. Note that each \EVS{} scan-line has its own color-coded energy axis. Dashed parts on the valley splitting traces indicate areas in which an anticrossing was not observable or out of $B$ range. (b) False-color 2D map of \EVS{} exclusively based on the data shown in panel a. (c) Correlation coefficient (dots) of the set of measured \EVS{} pairs separated by a geometric distance $D = \sqrt{\Delta y^2 + \Delta d^2}$, as a function of $D$, exclusively based on the data shown in panel a.  A Gaussian least-square fit to the correlation for $D<28$\,nm is included as a red solid line. (d) 2D map of \EST{} values obtained on the same wafer, but different device emplyoing magnetospectroscopy. \EST{} values are shown on the vertical axis as well as by the color of each bar. Green-blue stripes are the colored scanning electron micrograph of the clavier gates of the used device (cf. Fig.~\ref{fig:overview}a) (e) Histogram of the measured \EVS{} obtained by equidistant sampling of spline fits to the data of panel a (measured by coherent shuttling). (f) Histogram of the measured \EST{} using all data from panel d (measured by magnetospectroscopy). Both datasets are plotted with a maximum-likelihood fitted Rician distribution (solid line) and folded Gaussian distribution (black, dashed line).}
    \label{fig:Comparison}
\end{figure}
Hence, we could explain the features in Fig.~\ref{fig:Mapping}c, and found striking evidences that the green solid line in Fig.~\ref{fig:Mapping}d maps the \EVS{}($d$) along the 1DEC. The position along $B$ of this line can be resolved with a precision of less than \SI{1}{\micro \electronvolt} (see supplementary Fig.~S4). Care must be taken to interpret the plotted distance $d$ in terms of a precise location. $d(\tau_\mathrm{S})$ is extracted from the phase of the sinusoidal driving signal (Eq.~(\ref{eq: Shuttle pulse}) in the method section).
The travelling wave potential exhibits higher harmonics which leads to slight breathing and wobbling of the propagating QD, thus the QD velocity is not exactly constant. Slight variations in the velocity due to potential disorder from charged defects at the oxide interface are of the same order of magnitude \cite{Langrock23} imposing an uncertainty on QD position $d$. We note that we can shuttle the electron forth and back by a maximal one way distance of $d=\SI{336}{\nano \meter}$ equivalent to 1.2 $\lambda$. By reducing the shuttle velocity by a factor of five, we can shuttle the charge forth and back at at least 2.0 $\lambda$ ($d=\SI{480}{\nano \meter}$). This points to a potential-disorder peak at $d \! \approx \! \SI{340}{\nano \meter}$, which the electron cannot pass at the higher velocity. Here, we limit our mapping range to $d=\SI{210}{\nano \meter}$ (extended range shown in the supplementary material) to stay far away from this potential disorder peak, but also note that the abrupt change of $\nu$ and $\EVS{}$ at $d\approx 120$\,nm in Fig.~\ref{fig:Mapping}b,d indicates some tunneling occurring during the conveyor-mode shuttle process.

\addtocontents{toc}{\SkipTocEntry}
\subsection*{2D valley splitting map }

For simplicity, we approximate $d$ as the location of the QD now. In order to extend the mapping to the perpendicular direction, we change the screening gate voltages---from $V_\mathrm{ST}=V_\mathrm{SB}=\SI{100}{\milli \volt}$ while keeping the sum constant---in order to displace the 1DEC in the y-direction. Fig.~\ref{fig:Comparison}a displays the extracted splines corresponding to four different screening gate configurations where the nominal displacement in y-direction is indicated by colored labels. These distances are calculated by linearly converting the voltage difference $V_\mathrm{ST}-V_\mathrm{SB}$ into y-displacement with a factor of $\SI{6}{\nano \meter} / \SI{100}{\milli \volt}$ (see supplementary Fig.~S7e). The splines are sampled at the measurement resolution of one point per nominal \SI{1.4}{\nano \meter}. For some $d$ marked by dotted lines in Fig.~\ref{fig:Comparison}a (red trace:  $\sim$180-\SI{190}{\nano \meter}, violet trace: $\sim$170-\SI{185}{\nano \meter}, blue trace: $\sim$110-\SI{125}{\nano \meter}), we were unable to identify the \EVS{}, probably because it was below the $B$-scan range.

 Using all this data, we obtain a two-dimensional map of \EVS{} by linear interpolation (Fig.~\ref{fig:Comparison}b). The overall \EVS{} values are in the lower range of values found in the literature. The important point is, however, that our shuttling-based mapping method gives us an unprecedented insight into the lateral \EVS{} distribution in our SQS device. There are regions of nearly zero \EVS{} (e.g. $d\approx \SI{180}{\nano \meter}$ and $y=\SI{-12}{\nano \meter}$), but strikingly they can be avoided by displacing the QD along the $y$-direction (e.g. $y=\SI{6}{\nano \meter}$). This is important for shaping a static QD containing a spin-qubit at a position, at which \EVS{} is sufficiently large and qubit control is feasible. For conveyor-mode shuttling of spin-qubits, it allows finding a trajectory of the moving QD, which avoids low \EVS{} spots causing qubit decoherence. Similarly, tunneling of the moving QD across electrostatic disorder barriers (e.g. at $d\approx \SI{125}{\nano \meter}$ and $y=\SI{6}{\nano \meter}$) can be avoided by changing the y-displacement (e.g. $y=\SI{-12}{\nano \meter}$). The reason for the tuneability of \EVS{} is its short correlation length. 

We calculate the correlation coefficient of the set of \EVS{} pairs (without regions of undefined \EVS{}) separated by a geometric distance $D$ as a function of $D$ in Fig.~\ref{fig:Comparison}c. Additionally, we fit a Gaussian curve as derived from Ref.~\cite{Losert23}
\begin{equation}
    \mathrm{Corr}(D)=\exp\left(-\frac{1}{4-\pi} \frac{D^2}{a_\mathrm{dot}^2}\right),
    \label{eq:corr}
\end{equation}
which takes atomistic alloy disorder in the SiGe barrier into account. Here, the fitting parameter $a_\text{dot} = \hbar/\sqrt{m_t E_\text{orb}}$ is the characteristic QD size, $m_t$ is the transversal effective electron mass in silicon and $E_\text{orb}$ is the orbital energy of the electron, assuming a harmonic confinement potential.
The fit results in a QD size of $a_\mathrm{dot}\sim 16\, \mathrm{nm}$, corresponding to $E_\text{orb} \sim 1.6\,$meV being on the expected order of magnitude according to electrostatic simulations. Note that the correlation crosses zero and only vaguely follows a Gaussian decay, which is an effect of the limited scan area of the \EVS{} map (correlations of subsets of the data are discussed in the supplementary section S6). In addition, due to electrostatic disorder $E_\text{orb}$ is not constant, though assumed to be such in derivation of Eq.~(\ref{eq:corr}). 

\addtocontents{toc}{\SkipTocEntry}
\subsection*{Comparison to magnetospectroscopy}
In order to benchmark our new method for mapping the local \EVS{} by shuttling, we measure another map using the well-established method of magnetospectroscopy. We employ a device with the same heterostructure, gate-geometry and fabrication process, but the 1DEC is half in length and nine (instead of 17) individually tuneable (i.e. not interconnected) clavier gates are fabricated on top of the 1DEC (SEM is shown in the supplementary Fig.~S6a). We form a single QD at a time in the 1DEC by biasing some clavier gates and by the voltages $V_\mathrm{ST}, V_\mathrm{SB}$ applied to the long split-gate. To conduct the magnetospectroscopy, we tunnel-couple the QD to an accumulated electron reservoir reaching out to one SET, while the closer SET detects the charge-state of the QD (see supplementary section S7 for all details). We repeat the magnetospectroscopy each time forming a single QD at a different position in the 1DEC. The locations of these QDs (Fig.~\ref{fig:Comparison}d) are determined by triangulation with the QD's capacitive coupling to its four surrounding gates, and by a finite-element Poisson solver of the full device (see supplementary section S7). The orbital splitting $E_\mathrm{orb}$ of each QD is measured by pulsed gate spectroscopy yielding values in the range $E_\mathrm{orb}\sim 1.4 - 3.6\,$meV. By magnetospectroscopy, the two-electron singlet-triplet energy splitting \EST{} of the shaped QD can be directly measured. We nevertheless assume $\EVS{}\sim \EST{}$ to be a reasonable estimate, as the ratio between the two has been measured to be $E_\text{ST}/E_\text{VS}\lesssim 1$~\cite{Dodson22}, if $E_\mathrm{orb} \gg \EVS{}$ with \EVS{} then being weakly dependent on $E_\mathrm{orb}$~\cite{Ercan2021}.

This assumption allows comparing the histograms of both 2D maps (conveyor mode shuttling in Fig.~\ref{fig:Comparison}e and magnetospectroscopy in Fig.~\ref{fig:Comparison}f). Assuming that \EVS{} and \EST{} are both governed by alloy disorder, their distributions are expected to be Rician ~\cite{Losert23,Wuetz2022,delima23} 
\begin{equation}\label{eq:rice}
f(x  \, |  \, \gamma, \sigma) = \frac{x}{\sigma^2}  \exp\left(-\frac{x^2 + \gamma^2}{2\sigma^2}\right) I_0\left(\frac{x\gamma}{\sigma^2}\right).
\end{equation}
\noindent $I_0(x)$ is the modified Bessel function of the first kind and order zero.  $\gamma$ is the non-centrality parameter and $\sigma$ the scaling parameter. The fitted parameters $\gamma$ and $\sigma$ for both distributions (Tab. \ref{tab:histogram_fits}) are very similar. The $\sigma$ parameter expressing the randomness of the parameters is equal within the error range. The $\gamma$ parameter for \EST{} is a bit lower than the one of \EVS{} as expected. This all strongly supports the validity of our shuttle-based method for mapping the valley splittings. 
\begin{table}
    \centering
    \caption{Parameters fitting the distributions. In this table, we summarize the fit parameters that yield the fits in Fig. \ref{fig:Comparison}e for coherent shuttling (CS) and Fig. \ref{fig:Comparison}f for magnetospectroscopy (MS) with standard deviation $1\sigma$.}
    \begin{tabular}{|c|c|c|c|c|}
    \hline
    Parameter & value (CS) & $1\sigma$ (CS) & value (MS) & $1\sigma$ (MS) \\
    \hline
    \hline
    $\gamma$ & \SI{35.4}{\micro \electronvolt} & \SI{0.6}{\micro \electronvolt} & \SI{29.6}{\micro \electronvolt} & \SI{5.9}{\micro \electronvolt} \\
    \hline
    $\sigma$ & \SI{13.6}{\micro \electronvolt} & \SI{0.4}{\micro \electronvolt} & \SI{14.2}{\micro \electronvolt} & \SI{3.6}{\micro \electronvolt} \\
    \hline
    \hline
    $\mu$ & \SI{38.1}{\micro \electronvolt} & \SI{0.5}{\micro \electronvolt} & \SI{33.2}{\micro \electronvolt} & \SI{2.3}{\micro \electronvolt} \\
    \hline
    $\Tilde{\sigma}$ & \SI{13.0}{\micro \electronvolt} & \SI{0.3}{\micro \electronvolt} & \SI{13.1}{\micro \electronvolt} & \SI{1.8}{\micro \electronvolt} \\
    \hline
    \end{tabular}
    \label{tab:histogram_fits}
\end{table}

Intriguingly, we observe that $\gamma>\sigma$.  Consequently, both histograms can be well fitted by modified Gaussians (dashed lines in Fig. Fig.~\ref{fig:Comparison}e,f):
\begin{equation}
\begin{split}
    f(x \, |& \,  \mu, \Tilde{\sigma})=
   \frac{1}{ \sqrt{2\pi \tilde{\sigma}^2}} \times \\
   &\times \left(\exp{\left(-\frac{(x-\mu)^2}{2 \tilde{\sigma}^2} \right)} + \exp{\left(-\frac{(x+\mu)^2}{2 \tilde{\sigma}^2}\right)}\right),  
\end{split}
\end{equation}
\noindent with fitted parameters summarized in Tab. \ref{tab:histogram_fits}. This indicates that for both \EVS{} and \EST{} the randomness due to SiGe alloy disorder does not dominate over the deterministic contribution given by $\gamma$~\cite{Losert23,Wuetz2022,delima23}. However, care must be taken for the analysis of the histograms presented here, since a larger number of uncorrelated \EVS{} samples are required to reduce the error of the Gaussian tails. The samples for both histograms contain multiple points that are spatially closer than the fitted correlation length in Fig.~\ref{fig:Comparison}c. In addition, both histograms are slightly biased by omitting potentially a few small values due to the non-valid $\EVS{}(d)$ in Fig.~\ref{fig:Comparison}a. Especially, obtaining \EST{} smaller than the electron temperature by magnetospectroscopy is challenging and might explain that all \EST{}>\SI{12}{\micro \electronvolt}. In comparison, detecting \EVS{} lower than the electron temperature is possible by conveyor-mode shuttling. 

\addtocontents{toc}{\SkipTocEntry}
\section*{Discussion}
We introduced a new method for 2D mapping of the valley splitting \EVS{} in a Si/SiGe SQS with sub-\si{\micro \electronvolt} energy accuracy and nanometer lateral resolution. The method is based on separation and rejoining of spin-entangled electron pairs by conveyor-mode shuttling. Spin-singlet-triplet oscillations serve as a probe to identify spin-valley anticrossings and to extract the \EVS{} of both a static and a shuttled electron. The nanometer-fine tunability of the position of the shuttled QD allows for dense measurements, which allows us to identify local variations of the valley splitting landscape. By DC biasing the screening gates confining the 1DEC, we record a two-dimensional map of unprecedented large area. The method requires devices very similar to the ones used for quantum computation. Thus, the method is easily applicable and captures typical influences on the valley splitting e.g. effects from device fabrication. In principle, shuttling a single electron spin set in a spin-superposition is sufficient for our method. 

We benchmarked our results with magnetospectroscopy measurements---a well-established measurement method---on the same heterostructure and found the distributions of the measured map of singlet-triplet splittings to agree very well with the developed method. Note that mapping by magnetospectroscopy is limited in range due to the need of a proximate charge detector, and that the pure recording time required to obtain the presented 2D \EST{} map took us approximately 100 times longer than the more detailed \EVS{} map obtained by conveyor-mode shuttling.
While the extent of the latter map is spatially limited due to electrostatic disorder, we expect that higher confinement (large signal voltages) of the propagating QD will allow us to extend the mapped region. This new method offers a more comprehensive approach to heterostructure characterization and exploration, potentially aiding advancements in heterostructure growth and valley splitting engineering. Our results highlight the immediate benefits of conveyor-mode spin-coherent shuttling, not only for scaling up quantum computing systems but also for efficient material parameter analysis.

\addtocontents{toc}{\SkipTocEntry}
\section*{Methods}

\addtocontents{toc}{\SkipTocEntry}
\subsection*{Shuttle pulses}
In this section, we explain conveyor-mode electron shuttling in the 1DEC \cite{Struck23,Xue23,Seidler22,Kuenne23}. During the pulse stages T, $d$ and again T of the experiment, we apply sinusoidal pulses $V_{\mathrm{S},i}$ on the shuttle gates S$i$ (S1-S4):
\begin{equation}\label{eq: Shuttle pulse}
    V_{\mathrm{S},i}(\tau_\mathrm{S})=U_i\cdot\sin(2 \pi  f \tau_\mathrm{S}+\varphi_i)+C_i.
\end{equation}
The amplitudes ($U_1, U_3$) applied to the gate-sets $\mathrm{S1}$ and $\mathrm{S3}$ on the second layer (blue in Fig.~\ref{fig:overview}a) is $U_\mathrm{lower}=$\SI{150}{\milli \volt}, whereas the amplitudes ($U_1, U_3$) applied to the gate-sets $\mathrm{S2}$ and $\mathrm{S4}$ on the 3rd metal layer is slightly higher ($U_\mathrm{upper}=1.28\cdot U_\mathrm{lower}=$\SI{192}{\milli \volt}) to compensate for the difference of capacitive coupling of these layers to the quantum well \cite{Struck23}. This compensation extends to the DC-part of the shuttle gate voltages. The offsets $C_1=C_3= \SI{0.7}{\volt}$ are chosen to form a smooth DQD, whilst $C_2= C_4= \SI{0.896}{\volt}$ are chosen to form a smooth DC potential. The phases are chosen in order to build a travelling wave potential across the one-dimensional electron channel ($\varphi_1=-\pi/2, \varphi_2=0, \varphi_3=\pi/2, \varphi_4=\pi$) with wavelength $\lambda=\SI{280}{\nano \meter}$. The frequency $f$ is set to \SI{10}{\mega \hertz} resulting in a nominal shuttle velocity of \SI{2.8}{\meter \per \second}. The nominal shuttling distance $d$ relates to the assumption that the electron travels at a constant velocity $\lambda\cdot f$~\cite{Struck23}.

\addtocontents{toc}{\SkipTocEntry}
 \subsection*{Retuning SET and PSB}
In order to compensate for slow charge-noise drifts on the PSB and the SET, both the PSB-stage voltage as well as the SET voltages are retuned after 1000 repetitions. For this, we track the spin-fractions as well as the readout threshold between the charge configurations for singlet (4,0) and triplet (3,1). If we detect a significant change ($\sim 10\%$) in spin-fractions, this means the PSB region drifted and a correction via the G2 DC-voltage is done. Similarly, a significant change in readout threshold indicates a drift of the Coulomb-peak on the SET, resulting in need of adjusting its plunger voltage accordingly.

\addtocontents{toc}{\SkipTocEntry}
\subsection*{Layer stack of the used heterostructure}
The used Si/SiGe heterostructure is grown by chemical vapour deposition and has the following layer stack according to specification (top-to-bottom): Si-cap (\SI{2}{\nano \meter}), Si$_{0.7}$Ge$_{0.3}$ spacer (\SI{30}{\nano \meter}), strained Si quantum well (\SI{10}{\nano \meter}), Si$_{0.7}$Ge$_{0.3}$ barrier on virtual SiGe substrate.

\addtocontents{toc}{\SkipTocEntry}

\addtocontents{toc}{\SkipTocEntry}
\section*{Data availability}
The data that supports the findings of this study are available in the Zenodo repository \href{https://zenodo.org/doi/10.5281/zenodo.10359903}{(https://zenodo.org/doi/10.5281/zenodo.10359903)}.

\addtocontents{toc}{\SkipTocEntry}
\section*{Acknowledgements}
We acknowledge valuable discussions with Merritt P. Losert and Mark Friesen and the support of the Dresden High Magnetic Field Laboratory (HLD) at the Helmholtz-Zentrum Dresden - Rossendorf (HZDR), member of the European Magnetic Field Laboratory (EMFL). This work was funded by the German Research Foundation (DFG) within the project 421769186 (SCHR 1404/5-1) and under Germany's Excellence Strategy - Cluster of Excellence Matter and Light for Quantum Computing" (ML4Q) EXC 2004/1 - 390534769 and by the Federal Ministry of Education and Research under Contract No. FKZ: 13N14778, and by the National Science Centre (NCN), Poland under QuantERA program, Grant No. 2017/25/Z/ST3/03044. 
Project Si-QuBus received funding from the QuantERA ERA-NET Cofund in Quantum Technologies implemented within the European Union's Horizon 2020 Programme. The device fabrication has been done at HNF - Helmholtz Nano Facility, Research Center Juelich GmbH \cite{Albrecht17}.

\addtocontents{toc}{\SkipTocEntry}
\section*{Author contributions}
M.V., T.S. and B.C. set up and conducted the experiments assisted by L.V. Authors M.V., T.S., A.S., B.C., T.O., {\L}.C. and L.R.S. analysed the data supported by M.O. Device fabrication was done by J.T., R.X. and S.T. Author L.R.S. designed and supervised the experiment. L.R.S. and H.B. provided guidance to all authors. M.V., T.S., A.S. and L.R.S. wrote the manuscript which was commented on by all other authors.

\addtocontents{toc}{\SkipTocEntry}
\section*{Competing interests}
M.V., T.S., L.R.S. and H.B are co-inventors of patent applications that cover conveyor-mode shuttling and/or its applications. 
L.R.S. and H.B. are founders and shareholders of ARQUE Systems GmbH. The other authors declare no competing interest.


\clearpage
\onecolumngrid
\begin{center}
\textbf{\large Supplementary online material of\\Mapping of valley-splitting by conveyor-mode spin-coherent electron shuttling}
\end{center}

\vspace{1ex}
\begin{center}
Mats Volmer$^{1}$, Tom Struck$^{1,2}$, Arnau Sala$^{1}$, Bingjie Chen$^{1}$, Max Oberländer$^{1}$, Tobias Offermann$^{1}$, Ran Xue$^{1}$,
Lino Visser$^{1}$, Jhih-Sian Tu$^{3}$, Stefan Trellenkamp$^{3}$, {\L}ukasz Cywi{\'n}ski$^{4}$, Hendrik Bluhm$^{1,2}$, and Lars R. Schreiber$^{1,2}$
\\
\textit{\small $^{1}$JARA-FIT Institute for Quantum Information, Forschungszentrum\\
J\"ulich GmbH and RWTH Aachen University, Aachen, Germany\\
$^{2}$ARQUE Systems GmbH, 52074 Aachen, Germany\\
$^{3}$Helmholtz Nano Facility (HNF), Forschungszentrum J\"ulich, J\"ulich, Germany\\
$^{4}$Institute of Physics, Polish Academy of Sciences, Warsaw, Poland}
\end{center}
\normalfont

\setcounter{equation}{0}
\setcounter{section}{0}
\setcounter{figure}{0}
\setcounter{table}{0}
\setcounter{page}{1}
\makeatletter

\vspace{5ex}

\renewcommand{\thesection}{S\arabic{section}}
\renewcommand{\theequation}{S\arabic{equation}}
\renewcommand{\thefigure}{S\arabic{figure}}
\renewcommand{\thetable}{S\arabic{table}}

\twocolumngrid
\tableofcontents

\section{Anticrossing fitting algorithm}
In this section we describe the function that was used to fit the model from Eq.~(2) in the main text to the data. The approximated Hamiltonian in this equation is block-diagonal and hence could be diagonalized by hand. Nevertheless, in order to single out all the possible interactions that contribute to the formation of all the features observed in the data, we perform this analysis allowing for the possibility of including other Hamiltonian parameters, and thus resort to numerical diagonalization. For this, we define a function that takes as the argument the magnetic field as well as all parameters in Eq.~(2) ($\Delta g, E_r, E_l, v_r, v_l$). This function constructs and diagonalizes the Hamiltonian, and returns pairwise energy differences between all the combinations of eigenstates. We use these to calculate all the possible eigenfrequencies of the system and, then single out the smallest frequency, as the observed singlet-triplet precession frequency is governed by the smallest splitting in this model ($\Delta g \mu_B B \ll g \mu_B B, (E_l-E_r), E_r, E_l $).

\section{DQD valley splitting measurement at slanted screening gate voltages}
\begin{figure}
    \centering
    \includegraphics[width=\linewidth]{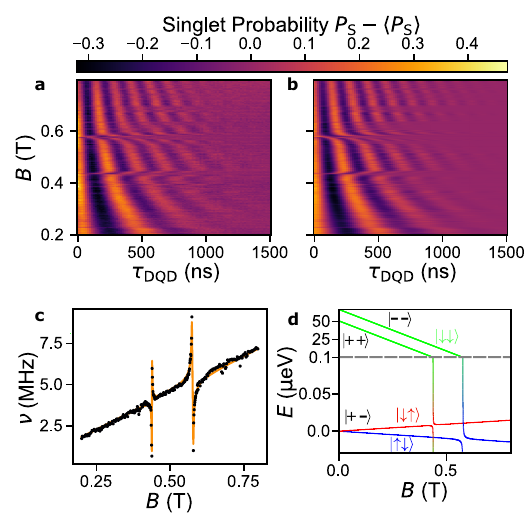}
    \caption{Spin-valley anti-crossing in a laterally shifted DQD at different y-displacement. The data corresponds to a similar experiment shown in Fig. 2 of the main.  Here, $V_\mathrm{ST}=\SI{50}{\milli \volt}$ and $V_\mathrm{SB}=\SI{150}{\milli \volt}$, yielding $y=\SI{-6}{\nano \meter}$. (a) Normalized singlet return probability $P_\mathrm{S}$ as a function of the magnetic field $B$ and DQD separation time $\tau_\text{DQD}$. $P_\mathrm{S}$ is normalized such that each horizontal line averages to zero. (b) Fit to the data from (a) using Eq.~(1) of the main text. (c) Frequencies $\nu$ extracted from the fit in (b). The orange curve is a least-square fit to the data. Uncertainties of frequencies are on the order of \SI{100}{\kilo \hertz} and smaller than the size of the black dots. (d) Energy spectrum of the Hamiltonian from main text Eq.~(2). The color mixture of the curves indicates the spin projection as indicated by the spin labeling (note the green-red and green-blue gradient near anticrossings), while the black symbols label the valley state. For clarity, the energy axis is upscaled around the states $\ket{\uparrow\downarrow +-}$ and $\ket{\downarrow\uparrow +-}$, with spin projection along the $z$ axis $m_\mathrm{S}=0$. For these states, their magnetic field dependence, proportional to $\Delta g \mu_B$, is four orders of magnitude smaller than that of the states $\ket{\downarrow\downarrow --}$ and $\ket{\downarrow\downarrow ++}$, with $m_\mathrm{S}=-1$. The parameters used in (d) are extracted from the fit in (c).}
    \label{Sfig:DQD_VS_slanted}
\end{figure}
The screening gate voltages used in Fig.~2 in the main text are $V_\mathrm{ST} = V_\mathrm{SB}=\SI{100}{\milli \volt}$. Here we report additional results for a different screening gate voltage configuration $V_\mathrm{ST}=\SI{50}{\milli \volt}$ and $V_\mathrm{SB}=\SI{150}{\milli \volt}$, yielding  $y=\SI{-6}{\nano \meter}$ (Fig. \ref{Sfig:DQD_VS_slanted}). The data is analyzed in the same way as in Fig. 2 from the main text. The corresponding fit parameters ($y=\SI{-6}{\nano \meter}$) as well as the fit parameters from the main text ($y=\SI{0}{\nano \meter}$) can be found in Tab. \ref{Stab:ac_fit_params}. The valley splitting as well as the spin-valley coupling of the right dot changes by a small margin.

\section{Time resolved anticrossing measurements in the 1DEC}
\begin{figure*}
    \centering
    \includegraphics[width=0.95\textwidth]{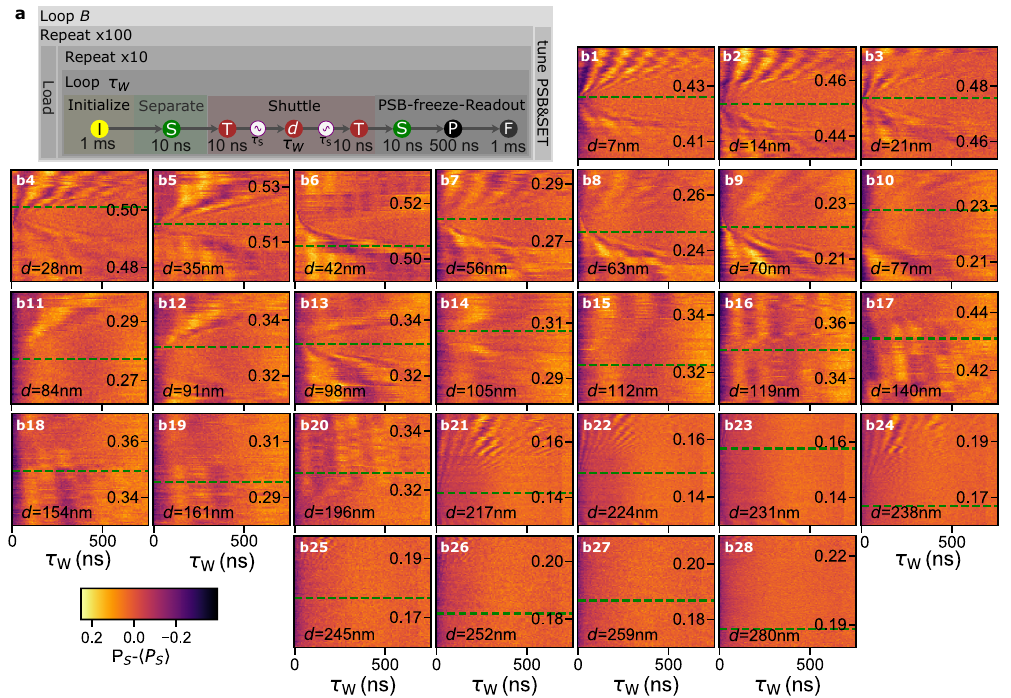}
    \caption{Time-resolved spin-valley anticrossing in the 1DEC. (a): Flowchart of the microscopic pulse stages for the experiment. The system is initialized in stage I, then one electron is separated and the tunnel barrier is pulled up (stages $\mathrm{S}\to\mathrm{T}$). The electron is shuttled at maximum speed for a distance $d$ where it is left to precess for varying $\tau_\mathrm{w}$ (stages $\mathrm{T}\to d$). Thereafter, the electron is shuttled back, the tunnel barrier is lowered and spin is converted to charge at the PSB (stages $d\to\mathrm{T}\to\mathrm{S}\to\mathrm{P}$). The charge state is frozen for readout (stages $\mathrm{P}\to\mathrm{F}$). (b1-b28): The panels show the spin-singlet probability as a function of evolution time $\tau_\mathrm{w}$ at  at the distance $d$ in the 1DEC given in the bottom left of the plot while varying the magnetic field. The position of the observed anticrossing in the corresponding valley splitting scan from the main text is indicated by a dashed green line for each $d$.}
    \label{Sfig:Anticrossing_in_channel}
\end{figure*}
We measured time resolved anticrossings inside the 1DEC by recording $P_\mathrm{S}(d,\tau_\mathrm{w}, B)$ (pulse scheme in Fig. \ref{Sfig:Anticrossing_in_channel}a), similar to the measurements from Fig.~2b of the main text, in which $P_\mathrm{S}(d,\tau_\mathrm{w}=\SI{300}{\nano \second}, B)$ is shown. The pulses schemes are equal, but we scan the singlet return-probability as a function of $\tau_\mathrm{w}$ for 28 different distances $d$ ranging from \SI{7}{\nano \meter} to \SI{280}{\nano \meter} with $\tau_\mathrm{w}$ ranging from 0 to \SI{700}{\nano \second}. After a coarse scan in $B$, we zoomed in the $B$ range of the spin-valley anticrossing (Fig. \ref{Sfig:Anticrossing_in_channel}b1-b28). The distance $d$ is inscribed in each panel in the bottom left. A horizontal green dashed line indicates the magnetic field, at which the anticrossing is observed in the corresponding $P_\mathrm{S}(d,\tau_\mathrm{w}=\SI{300}{\nano \second}, B)$ scan. The green-line appears in the $B$-region at which $\nu$ drastically changes. This confirms that we observe the spin-valley anticrossing in the Fig. 3 of the main text. For some $d$, the comparison is complicated by overlap of different valley splitting features. Minor deviations might originate from slightly different electrostatic tuning of the SQS as this measurement and the one presented in Fig. 3 of the main text have been conducted  with considerable time difference.

\begin{table}[]
    \centering
    \caption{Fit parameters, together with their uncertainty, for the model presented in Eq.~(2) of the main text, using the data from Fig.~\ref{Sfig:DQD_VS_slanted}b and d, corresponding to $y\! =\! -6$ nm (data for $y\! =\! 0$ from the main text are given for comparison). For the coupling elements $v_r$ and $v_l$, we also indicate the states that are coupled.}
    \begin{tabular}{|c|c|c|c|c|}
    \hline
        Parameter & value & $1 \sigma$ & unit/factor & coupling states \\
    \hline
    \hline
       $\Delta g \ (y=\SI{-6}{\nano \meter})$ & 6.46  & 0.01  & $10^{-4}$ & - \\
    \hline     
    
       $E_{r} \ (y=\SI{-6}{\nano \meter})$ & 50.74 & 0.03 & \si{\micro \electronvolt}& - \\
    \hline         
       $E_{l} \ (y=\SI{-6}{\nano \meter})$ & 66.74 & 0.02 & \si{\micro \electronvolt}& - \\
    \hline
        $v_r \ (y=\SI{-6}{\nano \meter})$ & 31 & 2 & neV & $\ket{\downarrow\uparrow +-}$ \& $\ket{\downarrow\downarrow ++}$ \\
    \hline
        $v_l \ (y=\SI{-6}{\nano \meter})$ & 68 & 1 & neV & $\ket{\uparrow\downarrow +-}$ \& $\ket{\downarrow\downarrow --}$ \\
    \hline
    \hline
       $\Delta g \ (y=\SI{0}{\nano \meter}) $ & 6.58& 0.04  & $10^{-4}$ & - \\
    \hline         
       $E_{r} \ (y=\SI{0}{\nano \meter})$ & 53.52 & 0.17 & \si{\micro \electronvolt}& - \\
    \hline         
       $E_{l} \ (y=\SI{0}{\nano \meter})$ & 66.64 & 0.04 & \si{\micro \electronvolt}& - \\
    \hline
        $v_r \ (y=\SI{0}{\nano \meter})$ & 82 & 14 & neV & $\ket{\downarrow\uparrow +-}$ \& $\ket{\downarrow\downarrow ++}$ \\
    \hline
        $v_l \ (y=\SI{0}{\nano \meter})$ & 58 & 3 & neV & $\ket{\uparrow\downarrow +-}$ \& $\ket{\downarrow\downarrow --}$ \\
    \hline
    \end{tabular}

    \label{Stab:ac_fit_params}
\end{table}

\section{Raw data from all valley splitting maps}
\begin{figure*}
    \centering
    \includegraphics[width=0.95\textwidth]{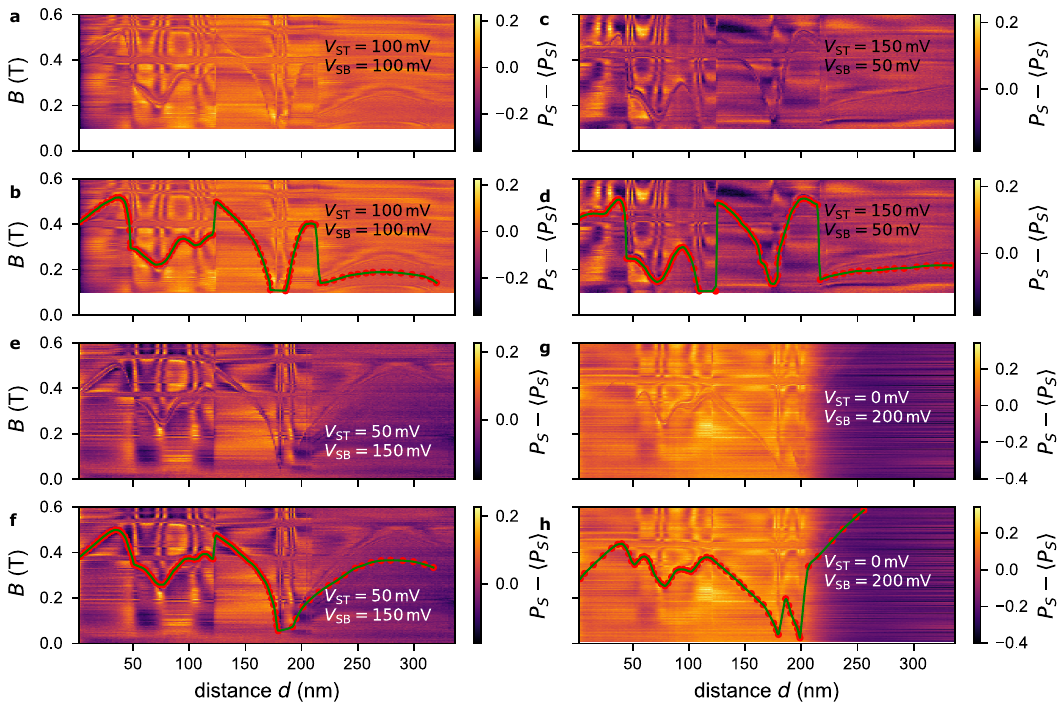}
    \caption{Mapping of the local valley splitting using the ST$_0$ oscillations for different $y$-displacements. (a, c, e, g): Singlet return probability $P_\mathrm{S}$ as a function of shuttle distance and magnetic field $d$. (b, d, f, h): Data from (a, c, e, g) with valley splittings extracted by hand (red dots) and a spline fit through the extracted points (green curve).}
    \label{Sfig:Raw_data_slanted_maps}
\end{figure*}
The rawdata and spline fits for all the four valley splitting scans $P_\mathrm{S}(d,\tau_\mathrm{w}=\SI{300}{\nano \second}, B)$ are shown in Fig.~\ref{Sfig:Raw_data_slanted_maps}. Each dataset is shown twice, with the raw singlet probability map on top and, the same map but with the spline fit as a guide to the eye below. We label each panel with the voltages applied to the two screening gates. The widths of the anticrossings become large and the mean $E_\text{VS}$ become constant at distances $d$ exceeding \SI{210}{\nano \meter}. As discussed in the main text, this might hint towards a disorder peak blocking the electron shuttling. Hence, we focus on data up to  $d=\SI{210}{\nano \meter}$ in the main text. We report our lowest measured valley splitting ($E_\mathrm{VS,min}\approx \SI{4.57}{\micro \electronvolt}$) in Fig. \ref{Sfig:Raw_data_slanted_maps}g,h at $d= \SI{200}{\nano \meter}$.

\newpage

\section{Energy resolution of the valley splitting maps}
\begin{figure*}
    \centering
    \includegraphics[width=0.95\textwidth]{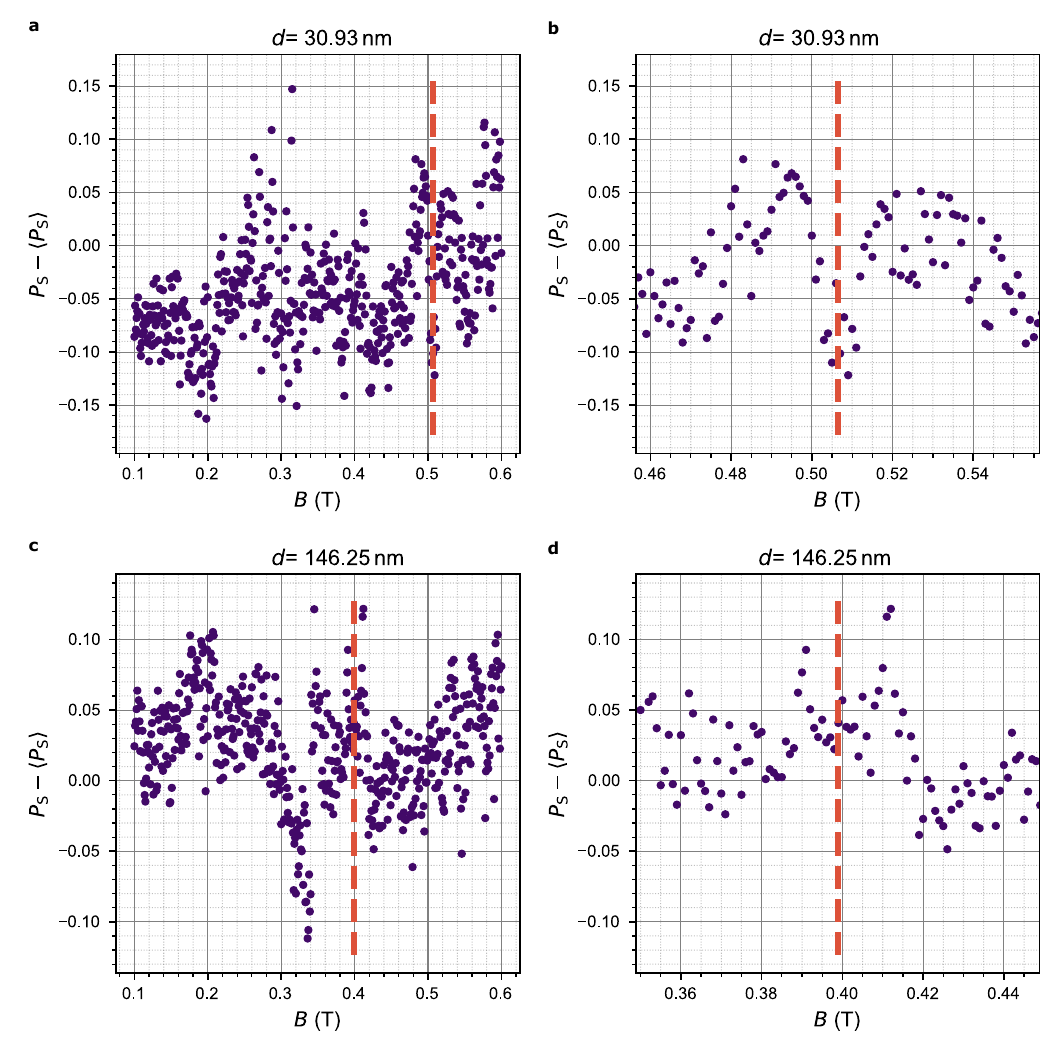}
    \caption{Linecuts through the valley splitting maps shown in Fig. 3c of the main text. (a) Full $B$-range linecut at $d=\SI{30.93}{\nano \meter}$, the vertical dashed line indicates the valley splitting as read off from Fig. 3d of the main text. (b) Zoom into a range of \SI{50}{\milli \tesla} around the valley-splitting. (c,d) Similar plots to (a,b), but for a linecut at $d=\SI{146.25}{\nano \meter}$.}
    \label{Sfig:dataslices}
\end{figure*}
We discuss how precise in $B$-field the energy of the anticrossing can be determined exemplary for Fig. 3c of the main text. Linecuts for exemplary two $d$ are shown in Fig.~\ref{Sfig:Raw_data_slanted_maps}. Obviously, the valley splitting cannot be determined from a single linecut as the signature of the anticrossing barley sets up from the background noise (Fig.~\ref{Sfig:Raw_data_slanted_maps}a,c). Only the contrast provided by the $P_\mathrm{S}(d,\tau_\mathrm{w}=\SI{300}{\nano \second}, B)$ map allows spotting and following the curved signature of the anticrossing. Therefore, we set the black dots in Fig. 3c of the main text by hand. Course identification of the anticrossing in terms of $B$-field, helps than to determine the valley splitting feature as a peak in the linecuts Fig.~\ref{Sfig:Raw_data_slanted_maps}b,d. We estimate that our readings have an inaccuracy of (less than) \SI{8}{\milli \tesla}, corresponding to an energy resolution of at least \SI{1}{\micro \electronvolt}.

\section{Correlation plot}
In Fig.~4 of the main text, we show the correlation coefficient as a function of a geometric distance using the data from Fig. 4a. In this section, we provide more details on the calculation of the correlation coefficient and extend the analysis by evaluating the correlation along the direction of the shuttling and the orthogonal one (y-direction).

\begin{figure}
    \centering
    \includegraphics[width = \linewidth]{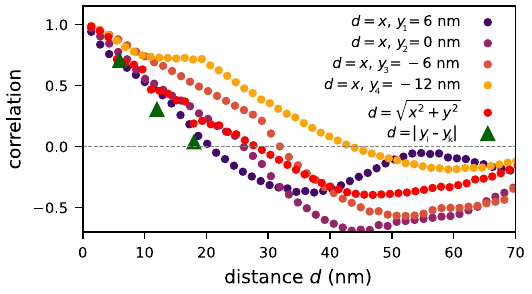}
    \caption{Correlation coefficient as a function of a distance. The points corresponding to $d=x$, for different $y_i$, show the correlation coefficient along the shuttling direction for each of the four different traces. The red dots show the correlation as a function of the geometric distance between pairs of data points belonging to the same or different traces. The green triangles show the correlation coefficient as a function of the distance along the direction orthogonal to the shuttling.}
    \label{Sfig:Supp_correlation.pdf}
\end{figure}

Each trace in Fig. 4a consists of a set of $\EVS$ measured at different positions along the direction of the shuttling. For each pair of data points, we measure the distance between them. We then sort all pairs in bins $\{\mathcal{B}_i\}$ such that each bin $\mathcal{B}_d$ contains all pairs of measured $\EVS$ separated by a distance $d$. Next, we calculate the Pearson's correlation coefficient for each of the bins. Repeating this process for each of the traces results in four colored sets of points in Fig.~\ref{Sfig:Supp_correlation.pdf}, labelled by corresponding values of shuttle path offset in the $y$ direction ($y=6,0,-6,12$\,nm).

The two-dimensional nature of our $\EVS$ map also allows for an evaluation of the correlation coefficient as a function of a geometric distance. For this, we redefine the distance $d$ as the geometric distance between a pair of data points from all the traces combined. Using this method, we obtain the red points in Fig.~\ref{Sfig:Supp_correlation.pdf}. This is the same correlation curve as shown in the main text. Finally, although we only have four traces along the direction orthogonal to the shuttling, we also evaluate the correlation coefficient along this direction, resulting in the three points (green triangles) in Fig.~\ref{Sfig:Supp_correlation.pdf}.
For the calculation of the correlation, we have omitted the points with unreliable information. These are indicated as dotted regions in the traces of Fig. 4a of the main text.
In all the cases, we see a correlation that decays to zero on a length scale of approximately $\sim 20 - 40$\,nm. As shown in the main text, this corresponds to an average QD size around $\sim 16$\,nm.

After the correlation curves cross the horizontal axis, we observe oscillations around zero. These are artifacts due to the short range of the data. We expect these oscillations to vanish in further experiments with a longer shuttling distance.

\section{Valley splitting mapping by magnetospectroscopy}
\begin{figure}
    \centering
    \includegraphics[width=\linewidth]{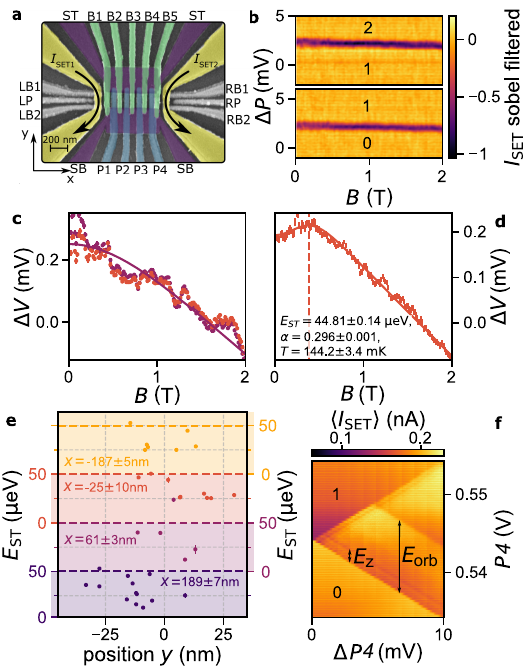}
    \caption{Magnetospectroscopy based valley splitting mapping on the same heterostructure. (a) SEM image of the measured device. (b) Sobel filtered raw data of the 01- and 12-transitions. The Sobel filter is applied in voltage sweep direction in order to better filter out the transition. (c) Fitted transition positions for the 01-transition (purple) as well as the 12-transition (red) with 1-$\sigma$ error bars. Also included is a fit through the 01-transition (solid line). (d) Noise-subtracted 12-transition position as a function of magnetic field with 1-$\sigma$ error bars on the plunger voltage. We extract the $E_\text{ST}$ by fitting Eq.~(\ref{eq:fit_EST}). (e) Mapping of the singlet-triplet splitting using the method described in (b-d) as well as the triangulation for 34 QD positions. The x axis represents the y-position on the sample. The measurements are grouped into four groups, each of them representing QDs under the four plungers (top to bottom in the plot corresponds to P1 through P4 in the sample). Inscribed are the x-positions of the QDs with calculated standard deviation. Error bars give the 1-$\sigma$-uncertainties. (f) Sample measurement of the orbital splitting via pulse spectroscopy. Here, we use a fifty percent duty-cycle pulse across the 01-Transition with varying amplitude (x-axis)  and offset (y-axis) to record the QD energy spectrum.}
    \label{Sfig:magnetospectroscopy}
\end{figure}

Here we provide more details about the valley splitting mapping using magnetospectroscopy presented in Fig.~4d,f in the main text. These measurements are conducted in a similar device on the same heterostructure. The gate design mainly differs in the length of the 1DEC, where the design used for the shuttling based measurements has seventeen clavier gates, and the one used for the magnetospectroscopy measurements has nine clavier gates (see Fig. \ref{Sfig:magnetospectroscopy}a). This design allows forming QDs by clavier and screening gates that have a well controllable tunnel coupling to a reservoir. The QDs are mainly confined under a plunger gate. The closest SET (for P1 and P2 the left SET; P3 and P4 the right SET) is used to perform charge readout. The reservoir is fed into the channel from the opposing side. We tune the QD in the reservoir into the few electron regime and perform magnetospectroscopy measurements \cite{SuppDodson22,SuppMcJunkin21}. Raw data is filtered by a Sobel filter, which is an image processing technique that detects edges in images by computing the gradient magnitude of image intensity. We use the kernel
\begin{equation}
    S_y=\left(\begin{array}{ccc}
        1 & 2 & 1 \\
        0 & 0 & 0 \\
        -1 & -2 & -1
    \end{array}\right), 
\end{equation}
to convolve the raw data. This effectively yields us a denoised version of the raw data, differentiated in voltage sweep direction. Lastly, we remove the SET-background of the voltage sweep by subtracting a median filter with a kernel that is three times the size of the transitions width. This leaves us with the data presented in Fig.~\ref{Sfig:magnetospectroscopy}b which is robust to fit. We fit the transition position using a Lorentzian peak and arrive at the data presented in Fig.~\ref{Sfig:magnetospectroscopy}c, which shows the position of the 01-transition (purple) and the 12-transition(red) as a function of the magnetic field with uncertainties. We then fit these points using the following expression~\cite{SuppMcJunkin21}:
\begin{equation}
    V_\mathrm{0 \to 1}(B)=-\frac{\beta B g \mu_B + 2 \log \left(e^{-\beta g \mu_B B} + 1 \right)}{2\alpha \beta} + V_0.
\end{equation}
Here $\beta=1/k_B T$ with $k_B$ being the Boltzmann constant and $T$ the temperature. $V_0$ is an offset fitting parameter and $\alpha$ the lever arm of the plunger on the QD. We see that slow noise with time constants on the order of the magnetic sweep speed deter both the 01 as well as the 12 transition. As these transition positions are recorded simultaneously, the presence of correlation in this noise is not surprising, and can be utilized to improve data quality significantly. We can use the 01-transition measurement as an effective noise measurement, and extract the noise on the transition by taking the residuals of the fit. Subtracting these residuals from the 12-transition, we arrive at the data shown in Fig.~\ref{Sfig:magnetospectroscopy}d. Now, a clearly visible kink emerges and can be fitted easily using the expression \cite{SuppMcJunkin21}
\begin{equation}\label{eq:fit_EST}
    V_\mathrm{1\to 2}(B) = \frac{1}{\alpha \beta}\log\left(\frac{(\epsilon+1)\epsilon^{\frac{1}{2}}e^{\beta E_\mathrm{ST}}}{\epsilon e^{\beta E_\mathrm{ST}} + \epsilon^2 + \epsilon + 1}\right)+V_0,
\end{equation}
with $\epsilon=e^{\beta g \mu_B B}$. From this fit we extract the singlet-triplet splitting $E_\mathrm{ST}$. After this measurement, we estimate the position of the QD by triangulation, as described below. We repeat all these steps for a set of different screening gate voltage configurations for QDs under all four plungers, and arrive at the mapping shown in Fig. \ref{Sfig:magnetospectroscopy}e. The data displayed here is the same as in the main text Fig. 4c, only shown in a more analytical way, with more focus on the quantitative valley splitting values. Here, the x-axis shows the y-position of each measurement. The y-axis shows the measured singlet-triplet splitting. The measured points are grouped into one group for each plunger. The respective average x-position with standard deviation is inset in the plot.

Lastly, in Fig. \ref{Sfig:magnetospectroscopy}f we show a sample pulse spectroscopy measurement of the orbital splitting. By using the lever arm evaluated by magnetospectroscopy, we can translate the observed line to an orbital splitting (in this case \SI{2.3}{\milli \electronvolt}). The additional downwards sloping line at low energies corresponds to the Zeeman splitting, as the magnetic field of this measurement is \SI{3}{\tesla}.
\begin{figure}
    \centering
    \includegraphics[width=7.5 cm]{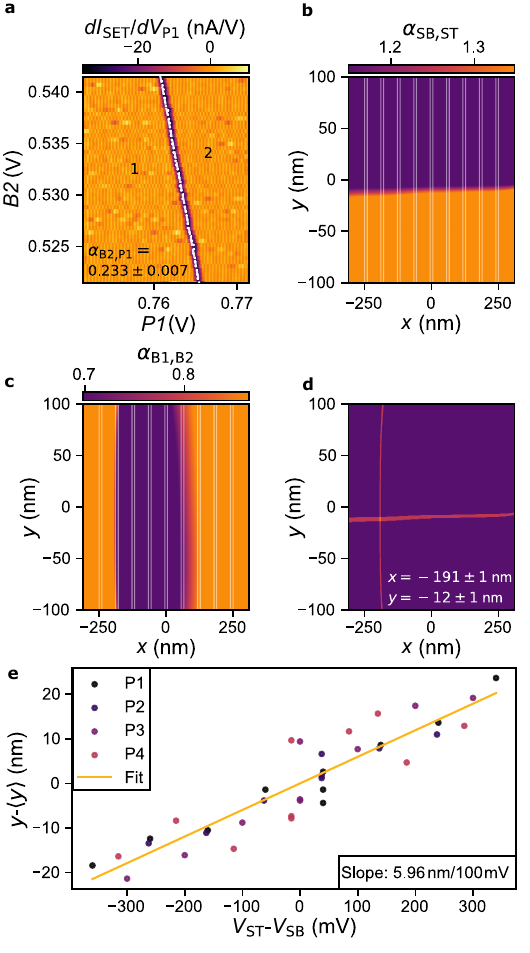}
    \caption{Determination the QD's coordinates for the valley-splitting maps. (a-d) Triangulation method applied to each QD formed during magnetospectroscopy. Here, we illustrate the method for one exemplary QD formed underneath plunger gate P1. (a) Charge stability diagram recorded by the left SET as a function of the voltages applied to the right barrier gate B2 and the plunger gate P1. The dashed line represents a least-square fit to the 1,2-charge transition line. (b) Exemplary simulated ratio of cross-capacitances as a function of QD ($x,y$)-coordinates for variation on the screening gates SB and ST.  The contrast of the color bar is adjusted to the measured cross-capacitance ratio $\alpha_\mathrm{SB,ST}=1.25\pm0.10$. (c) Same as in panel b, but for the ratio of cross-capacitances of the left (here B1) and right barrier gates (here B2) to the QD (measured cross-capacitance ratio $\alpha_\mathrm{B1,B2}=0.78\pm0.08$). (d) Triangulation overlay of results from panel b and c with $1\sigma$ on each cross-capacitance (represented by line-width). The QD position is determined by the intersection of the two lines (yellow area). (e) Evaluation of the average y-displacement as a function of the difference of voltages applied to the screening gates ST and SB ($V_\mathrm{ST}-V_\mathrm{SB}$). Dots represent the y-position of all QDs determined by our triangulation method. The dot-color indicates the adjacent plunger gate of the formed QD, which was also used for the triangulation method. The y-positions are plotted relative to the average of all y-coordinates. Slope of least-square linear fit (yellow line) gives the the average y-displacement as a function of $V_\mathrm{ST}-V_\mathrm{SB}$.}
    \label{Sfig:Triangulation}
\end{figure}

The (x,y)-coordinates of each QD formed for mapping by magnetospectroscopy is determined by measurement and simulation of cross-capacitances of the QD with proximate gates. First, we measure the ratio of the cross-capacitance of the two screening gates ST and SB and the ratio of the cross-capacitance of the two adjacent barrier gates LB (left barrier) and RB (right barrier). Second, we compare the results to corresponding cross-capacitances ratios simulated by a finite-element Poission solver of the full device. Since gates ST and SB are perpendicular to gates LB and RB by design, triangulation of the QD position is easy. In addition, the measurement of all cross-capacitances takes some random shift of the QD position due to electrostatic disorder into account.

In the following, we explain the triangulation for one exemplary QD, here formed close to the plunger gate P1. After the magnetospectroscopy measurement, we record four charge stability diagrams as a function of the voltages applied to the plunger gate (P) sitting on top of the formed QD and of the voltage applied to one the four gates LB, RB, ST and SB. Each stability diagram is measured  at the 1-2 charge transition, thus at the operation point of the magnetospectroscopy (see Fig.~\ref{Sfig:Triangulation}a with gates LB, P represented by B1 and P1, respectively). From this, we determine the ratios of cross-capacitances between both the adjacent barriers and the QD $\alpha_\mathrm{LB,RB}$ as well as the ratio of cross-capacitance between the both screening gates and the QD $\alpha_\mathrm{ST,SB}$ with
\begin{equation}\label{eq:ccs}
\alpha_\mathrm{g_\mathrm{1},g_\mathrm{2}}=\frac{\alpha_\mathrm{g_\mathrm{1},P}}{\alpha_\mathrm{g_\mathrm{2},P}},
\end{equation}
where $\alpha_\mathrm{g_\mathrm{1},g_\mathrm{2}}$ is the ratio of cross-capacitance of the gates $\mathrm{g_\mathrm{1}}$ and $\mathrm{g_\mathrm{2}}$ to the QD.   

Next, we simulate the electrostatics of the full, ideal device including all layers and gates, but excluding any sources of electrostatic disorder by COMSOL Multiphysics\textsuperscript{\textregistered} \cite{SuppCOMSOL} finite-element Poisson solver. We simulate the ratios of cross-capacitance between LB and RB to the QD $\alpha_\mathrm{LB,RB}(x,y)$ as well as SB and ST to the QD $\alpha_\mathrm{SB,ST}(x,y)$ for various $(x,y)$ coordinates of the QD. To this end, we use the DC voltages applied on all gates ($V_\mathrm{op}$) in the experiment as input parameters and vary voltages by $\Delta V_\mathrm{gate}$ around this operation point:
\begin{equation}\label{eq:diff_quot}
\begin{split}
    \alpha_\mathrm{g1,g2}&(x,y)=  \\
    & \frac{U_\mathrm{el}(V_\mathrm{op}+\Delta V_\mathrm{g1};x,y)-U_\mathrm{el}(V_\mathrm{op}-\Delta V_\mathrm{g1};x,y)}{U_\mathrm{el}(V_\mathrm{op}+\Delta V_\mathrm{g2};x,y)-U_\mathrm{el}(V_\mathrm{op}-\Delta V_\mathrm{g2};x,y)},
\end{split}
\end{equation}
where $U_\mathrm{el}(V_\mathrm{op}+\Delta V_\mathrm{gate};x,y)$ is the electrostatic potential at $V_\mathrm{op}$, with an added small voltage $\Delta V_\mathrm{gate}=\SI{5}{\milli \volt}$ on the gate as a function of ($x,y$). Calculating this for all the adjacent gates, we evaluate the spatially dependent cross-capacitances $\alpha_\mathrm{SB, ST}(x,y)$ (Fig. \ref{Sfig:Triangulation}b) and $\alpha_\mathrm{LB,RB}(x,y)$ (Fig. \ref{Sfig:Triangulation}d with gates LB, RB represented by B1 and B2, respectively). In the last step, we compare the measured values $\alpha_\mathrm{SB,ST}$ and  $\alpha_\mathrm{LB,RB}$ to all simulated cross-capacitance ratios, and clip the simulated values to the measured value within a vicinity given by $1\sigma$ uncertainty of the measurement. Finally, we overlay both clipped areas to determine the QD's position (Fig. \ref{Sfig:Triangulation}d), where the uncertainty of ($x,y$) is given by the standard deviation of the overlap. We repeat the triangulation procedure for all QDs formed for the magnetospectroscopy measurement.

For the valley-splitting map in Fig. 4a and b of the main text, we require the $y$-displacement as a function of the difference of voltages applied to the screening gates ST and SB ($V_\mathrm{ST}-V_\mathrm{SB}$). We determine the influence of the screening gates ST and SB by averaging the QD $y$-positions determined from the triangulation measurements. Hence, the electrostatic disorder in the 1DEC enters this average to some extent. We find that the averaged y-displacement is $\approx \SI{6}{\nano \meter}/\SI{100}{\milli\volt}$ per voltage on the screening gates (Fig. \ref{Sfig:Triangulation}e).

\newpage

\let\oldaddcontentsline\addcontentsline
\renewcommand{\addcontentsline}[3]{}

\let\addcontentsline\oldaddcontentsline

\end{document}